\documentclass{aa}  

\usepackage{graphicx}
\usepackage{txfonts}
\newcommand {\ixpe}{\text{IXPE}\xspace}
\newcommand {\swift}{\textit{Swift}/XRT\xspace}
\newcommand {\xmm}{\textit{XMM--Newton}\xspace}

\newcommand {\source}{{GX\,13$+$1}\xspace}

\begin{document}

   \title{New polarimetric study of the galactic X-ray burster \source}

    \author{
        Anna Bobrikova \inst{\ref{in:UTU}}
        \and Alessandro Di Marco \inst{\ref{in:INAF-IAPS}}
        \and Fabio La Monaca \inst{\ref{in:INAF-IAPS},\ref{in:UniRoma2},\ref{in:LaSapienza}} 
        \and Juri Poutanen \inst{\ref{in:UTU}} 
        \and Sofia V. Forsblom \inst{\ref{in:UTU}}
        \and Vladislav Loktev \inst{\ref{in:UTU}}}

   \institute{
        Department of Physics and Astronomy, FI-20014 University of Turku,  Finland \label{in:UTU} \\ 
        \email{anna.a.bobrikova@utu.fi} 
        \and 
        INAF Istituto di Astrofisica e Planetologia Spaziali, Via del Fosso del Cavaliere 100, 00133 Roma, Italy \label{in:INAF-IAPS}
        \and
        Dipartimento di Fisica, Universit\`{a} degli Studi di Roma ``Tor Vergata'', Via della Ricerca Scientifica 1, 00133 Roma, Italy \label{in:UniRoma2} 
        \and 
        Dipartimento di Fisica, Universit\`{a} degli Studi di Roma ``La Sapienza'', Piazzale Aldo Moro 5, 00185 Roma, Italy \label{in:LaSapienza}}
   \date{Received April 2, 2024; accepted XXX}

 
\abstract{Weakly magnetized neutron stars (WMNSs) are complex astrophysical objects with challenging phenomenology. For decades, they have been studied via spectrometry and timing analyses. It is well established that the spectrum of WMNSs consists of several components traditionally associated with the accretion disk, the boundary or spreading layer, and the wind, along with their interactions. Since 2022, WMNSs have been actively observed using the Imaging X-ray Polarimetry Explorer (\ixpe). Polarimetric studies have provided new information about the behavior and geometry of these sources.
One of the most enigmatic sources in this category, the galactic X-ray burster \source, was first observed with \ixpe in October 2023. 
A highly variable polarization at levels of 2--5\% was detected, with the source showing a rotation of the polarization angle (PA), suggestive of misalignment within the system. 
A second observation was performed in February 2024, complemented by observations from \swift. 
\ixpe measured an overall polarization degree (PD) of 2.5\% and a PA of 24\degr, while \swift data helped us evaluate the galactic absorption and fit the continuum. 
Here, we study the similarities and differences in the polarimetric properties of the source during the two observations. Our findings confirm the expected misalignment in the system and the assignment of the harder component to the boundary layer. 
We also emphasize the significance of the wind in the system. 
Additionally, we observe notable differences in the variation of polarimetric properties with energy and over time. }

\keywords{accretion, accretion disks – polarization – stars: neutron – X-rays: binaries
               }

   \maketitle
%

\section{Introduction}

Although weakly magnetized neutron stars (WMNSs) do not show pulsation or strong orbital variation, they present a complex phenomenology that requires further studies and observations to be explained. The broad class of WMNSs includes sources with a magnetic field on the order of $10^7-10^9$ G. These neutron stars accrete matter from their companions via the Roche lobe. If the accretion rate is high enough to overpower the relatively weak magnetic field, the matter from the accretion disk falls directly onto the equator of the neutron star. 

Non-pulsating WMNSs are traditionally classified into two main groups, Z and atolls, based on the patterns they show on the color-color diagram (CCD) or hardness-intensity diagram (HID).  
Although the classification is completely phenomenological, objects of these two classes have many other characteristic behavioral properties. For instance, atolls are known to be less bright than the Z sources: the typical luminosity of the atolls is in the range of $10^{36}-10^{37}$~erg~s$^{-1}$, while Z sources are among the brightest objects in the X-ray sky, with luminosities on the order of $10^{38}$~erg~s$^{-1}$.
However, many features are known to be present in both classes. For example, quasi-periodic oscillations (QPOs) in the hertz and kilohertz ranges were observed from both Z and atoll sources \citep{vanderklis89,vanderklis00}. Sources of both classes are observed in both hard and soft states, and the main components of their spectrum are expected to be the same. The spectrum is generally well described by the two main components \citep[see, e.g.,][]{Revnivtsev13}. The soft component can be approximated by the multicolor blackbody emission of temperature $<1$~keV associated with the accretion disk. The harder component is probably caused by Comptonization in a relatively cool plasma (2--3~keV), either a boundary layer between the accretion disk and the surface of the neutron star \citep{Shakura88} or the spreading layer at the neutron star surface \citep{inogamov1999,suleimanov2006}. Additional features known to the spectrum are the broad Fe emission line associated with the reflection of the spreading layer emission from the accretion disk and sharp absorption lines caused by the absorption of the emission in the wind above the disk. 

However, the geometry of the WMNSs cannot be fully understood from spectroscopic and time-resolved observations alone. 
Information obtained from variability, and specifically from the QPOs, revealed that the strongest variability is associated with a harder spectral component \citep{gilfanov2003, Revnivtsev06, Revnivtsev13} originating close to the neutron star surface. 
Polarimetry brought a new perspective to the problem. For instance, it became possible to distinguish between the boundary and spreading layer emission based on the polarimetric properties of the emission, while spectroscopically, they are identical. The boundary layer is expected to have a polarization similar to that of the accretion disk, with the polarization vector lying in the plane of the disk \citep{Chandrasekhar1960, Loktev22}, while the polarization of the spreading layer is expected to be perpendicular to the accretion disk \citep{Farinelli24}. 
For example, the polarization angle (PA) of the bright Z-source \mbox{Cyg X-2} \citep{Farinelli23} turned out to be consistent with the position angle of the radio jet (which is likely perpendicular to the accretion disk), implying the polarization originated in the spreading layer or from scattering in the accretion disk wind (\citealt{2024MNRAS.527.7047T}; Nitindala et al., in prep.). 
In \citet{LaMonaca2024}, on the other hand, a significant difference between the jet position angle and the PA for \mbox{Sco X-1} has been observed, which implies a more complicated geometry of the source. 
In the peculiar case of \mbox{Cir X-1} \citep{Rankin2024}, the variability of the PA with time and hardness supports the idea of a misalignment between the angular momentum of the neutron star and the binary orbital axis. 
In Z-sources \mbox{XTE J1701$-$462} \citep{Cocchi2023} and \mbox{GX 5$-$1} \citep{Fabiani24}, a significant variation in the polarization between observations in the hard and soft states suggested a change in geometry between states. 
In atoll sources \mbox{GX 9+9} \citep{Ursini23}, \mbox{4U 1820$-$303} \citep{DiMarco23}, and \mbox{4U 1624$-$49} \citep{Saade24}, the strong dependence of the polarization degree (PD) on energy supports the separation of the spectrum into two components with the different PAs. 
These discoveries demonstrate the usefulness of using polarimetric observations to study these sources. 

The galactic X-ray burster \source was first observed by Imaging X-ray Polarimetry Explorer (\ixpe) on 2023 October 17--19 \citep[][hereafter B24]{Bobrikova24}. This source is known for the strong wind above the large disk being responsible for several absorption lines \citep{DT12}, a brightness of $0.5 L_{\rm Edd}$ \citep{Dai14}, and an orbital period of 24.5~d, which is very long for a low-mass X-ray binary \citep{Corbet10}. \source presents a classification challenge: it is still debatable whether it is a Z \citep[see, e.g.,][]{Saavedra23} or an atoll source \citep[see, e.g.,][]{Schnerr03}. 

During the first \ixpe observation, \source presented an enigmatic behavior: the polarimetric properties varied significantly during the observation, while the spectral properties remained almost unchanged. Moreover, \source shows a peculiar continuous rotation of the PA by 70\degr\ in the two days of observations, together with a change in the pattern of dependence of PD on energy. These two peculiar behaviors left some space for further investigation. 
Here, we aim to answer some of the open questions and bring a new perspective to the study of \source. 

The remainder of the paper is structured as follows: we introduce the observations performed with the \ixpe and \swift observatories in Sect.~\ref{sec:obs}. 
In Sect.~\ref{sec:analysis}, we present our data analysis. We provide possible interpretations of the results in Sect.~\ref{sec:discussion} and a summary in Sect.~\ref{sec:summary}.

\section{Observations}\label{sec:obs}

\subsection{IXPE} 

\ixpe is a joint mission of the National Aeronautics and Space Administration (NASA) and the Italian Space Agency (Agenzia Spaziale Italiana, ASI) that was launched on 2021 December~9. 
The observatory comprises three grazing-incidence mirror assembly modules, each of which has a polarization-sensitive X-ray detector unit (DU)  in the focal plane  \citep{Costa2001, Baldini21}. A description of the mission and of the instrument on board is given in \citet{Soffitta21} and \citet{Weisskopf2022}. 

\begin{figure}
\centering
\includegraphics[width=0.5\textwidth]{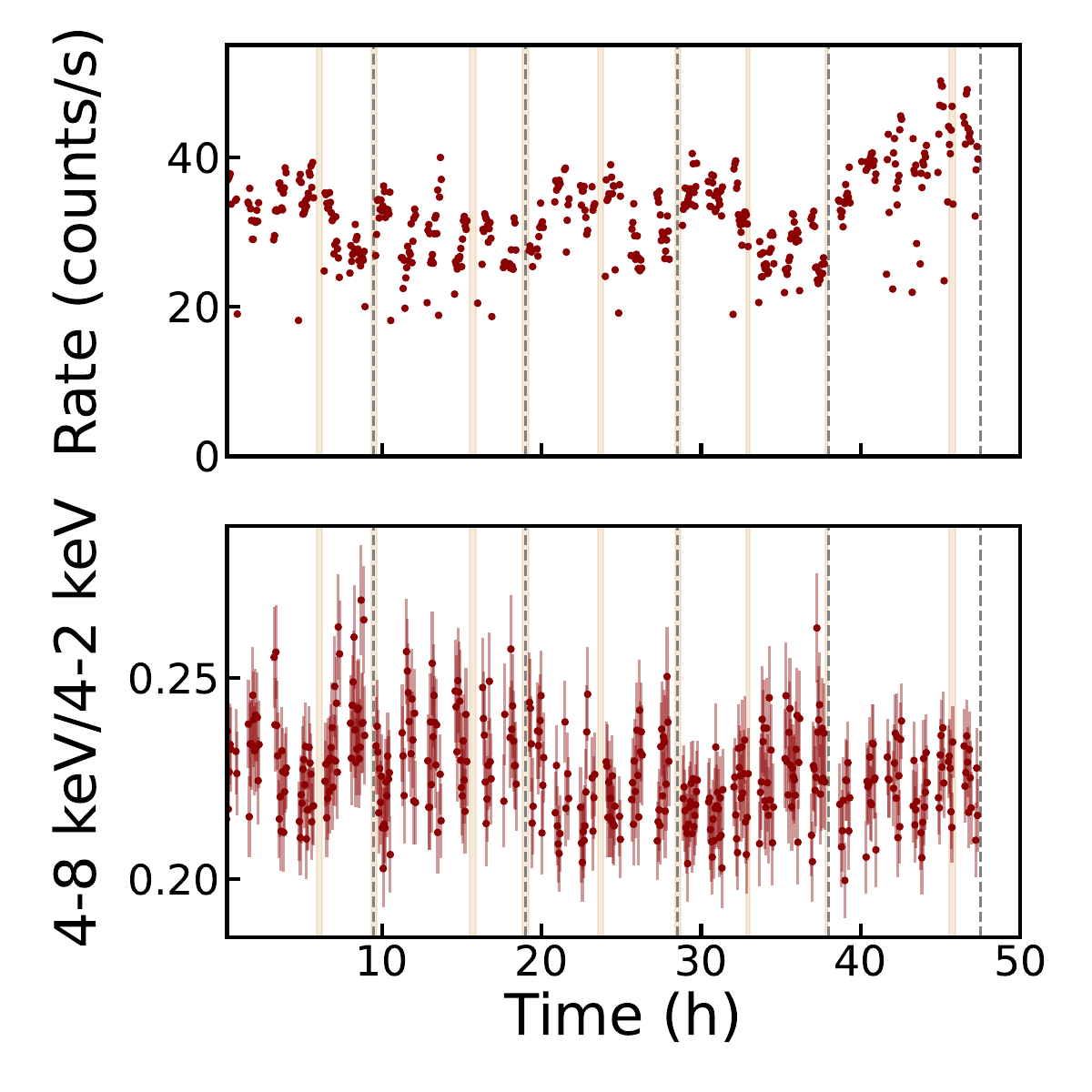} 
\caption{ \ixpe light curve obtained by combining the three DUs (top panel) and the corresponding hardness ratio as a function of time (bottom panel). The data are binned in 200~s. Shaded regions indicate the \swift observation periods. Dashed vertical lines indicate the separation of the observation into five equal 9.5~h time bins.}
\label{fig:lc}
\end{figure}

\ixpe observed \source twice: the first observation was conducted in October 2023, as reported in B24, and the second took place from 2024 February 25, at 13:34 UTC to 2024 February 27, at 12:59 UTC, with a total exposure time of approximately 90~ks for each DU (see Table~\ref{tab:observations} and the light curve in Fig.~\ref{fig:lc}).
Data were processed for model-independent polarimetric analysis using the \textsc{ixpeobssim} package version 30.2.1 \citep{Baldini2022}. Spectral and spectropolarimetric analysis was performed using \textsc{HEASoft} version 6.33 and the standard \textsc{FTOOLS} \citep{heasoft}, with the \ixpe Calibration Database (CALDB) that was released on 2024 February 28. Source photons were selected in a circular region with a radius of 100\arcsec\ centered at the source position. As recommended by \citet{DiMarco_2023}, given the brightness of the source, the background was not subtracted. The unweighted analysis was performed with \textsc{ixpeobssim}, while the weighted analysis was adopted for the spectropolarimetric analysis, as suggested by \citet{DiMarco_2022}. 

\begin{table*}
\centering
\caption{List of \ixpe and \swift observations.}
\label{tab:observations}
\begin{tabular}{lcccc}
\hline\hline
{Observatory} & {ObsID} & {Start -- Stop} & {Instrument} & {Exposure Time (s)} \\
            \hline
\textit{Swift} & 00036688042 & {2024-02-26T05:05}--\text{2024-02-26T05:27} & XRT & 1366 \\
& 00036688043 & \text{2024-02-25T13:14}--\text{2024-02-25T17:59} & & 892 \\
& 00036688044 & {2024-02-26T13:07}--{2024-02-26T14:52}& & 1249 \\
& 00036688045 & {2024-02-25T22:53}--{2024-02-26T00:35} & & 1395 \\
 & 00036688046 & {2024-02-26T17:56}--{2024-02-26T19:30} & & 1311 \\
 & 00036688047 & {2024-02-26T08:22}--{2024-02-26T10:12} & & 1429 \\
& 00036688048 & {2024-02-25T19:28}--{2024-02-25T21:32} & & 1321 \\
& 00036688049 & {2024-02-26T22:24}--{2024-02-26T23:59} & & 771 \\
& 00036688050 & {2024-02-27T03:22}--{2024-02-27T05:14} & & 507 \\
& 00036688051 & {2024-02-27T11:05}--{2024-02-27T11:29} & & 1443 \\
\hline
\ixpe & 03001101 & {2024-02-25T13:34}--{2024-02-27T12:59} & {DU1} & 90670 \\ 
& & & {DU2} & 90819 \\
& & & {DU3} & 90799 \\
\hline
\end{tabular}
\end{table*}

\begin{figure}
\centering
\includegraphics[width=0.5\textwidth]{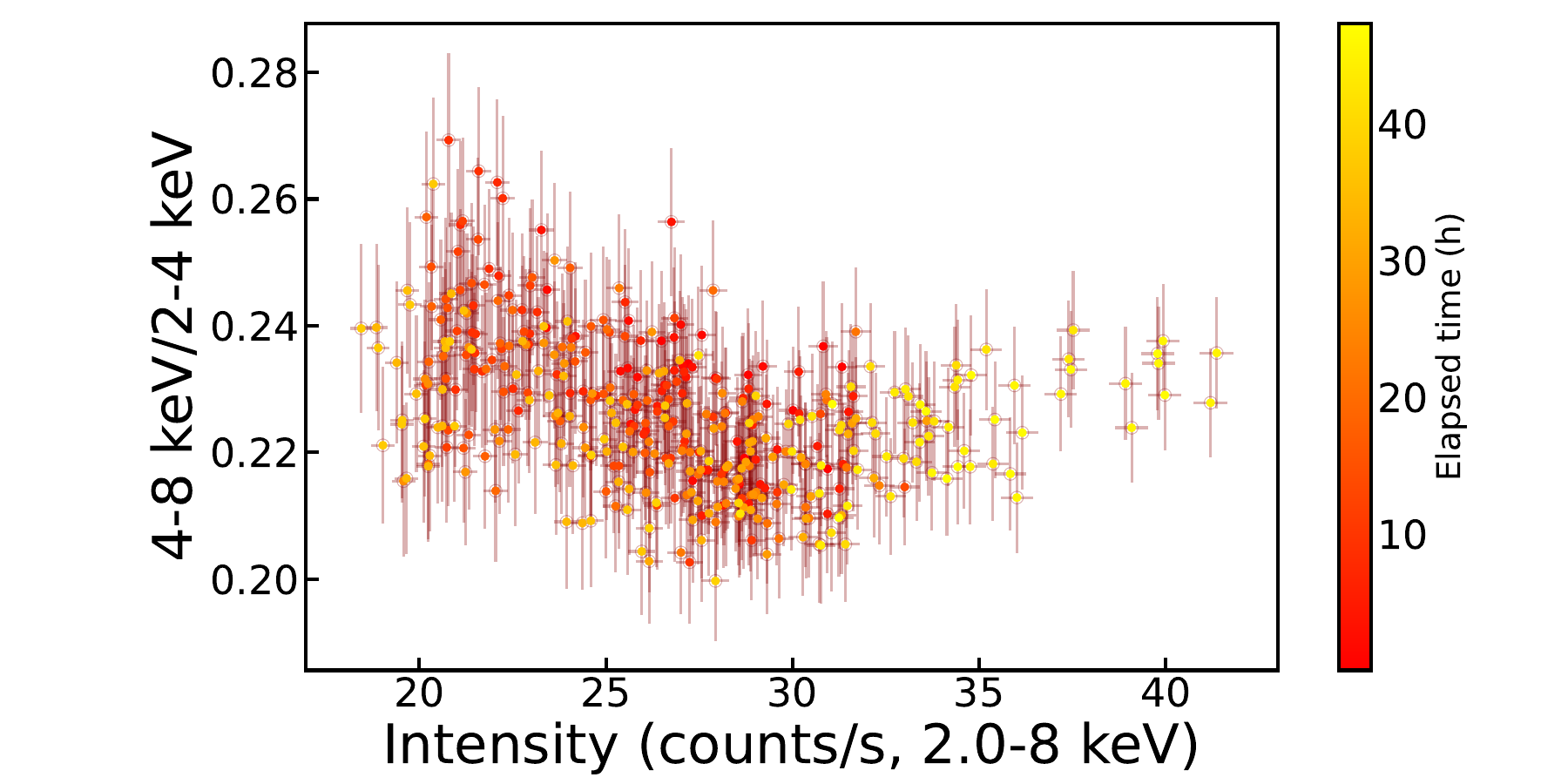}
\caption{Hardness-intensity diagram obtained from \ixpe data. 
The data are binned in 200~s.}
\label{fig:hid}
\end{figure}

\subsection{\swift}

Carrying three instruments on board, the \textit{Neil Gehrels Swift} Observatory \citep{Gehrels2004} enables the most detailed observations of gamma-ray bursts to date; the X-ray Telescope (XRT), one of the instruments on board, is based on a sensitive, flexible, autonomous X-ray CCD imaging spectrometer. 
\swift-coordinated observations with \ixpe for \source were performed, and used to monitor the status of the source and obtain spectral information during \ixpe observations. 
Given the source brightness, the \swift observations were performed in windowed timing (WT) mode. Ten \swift pointings covered the \ixpe observation, as reported in Table~\ref{tab:observations} and Fig.~\ref{fig:lc}. 
\swift data were extracted using \textsc{HEASoft} v6.33 and standard \textsc{ftools} \citep{heasoft}. 
Source and background extractions were performed using \swift imaging capabilities in an annulus with an inner radius of about 5\arcsec\ and an outer radius of about 60\arcsec\ to avoid possible pile-up effects \citep{Romano2006}. 
The \swift data were fitted in the energy band 0.7--8~keV to ensure sufficient statistics in the spectral fits and grouped to have at least 50 counts bin$^{-1}$. 
Response matrices released in the HEASARC CALDB on 2023 July 25, were applied in this analysis.

\section{Data analysis}\label{sec:analysis}

\ixpe observed \source for $\sim$90~ks. During the observation, as shown in Fig.~\ref{fig:lc} (top panel), the source rate was rather stable, with a slight increase at the end of the observation. The source hardness in the \ixpe data appears to have no strong variations, showing only a slight decrease with the count rate increase toward the end of the observation; this contrasts with the previous \ixpe observation reported in B24, where a dip corresponding to a slight hardening was present. 
Figure~\ref{fig:hid}, which reports the HID obtained from the \ixpe data, shows that hardness did not change by more than 10\%.

Consisting of short snapshots during the IXPE observation, the \swift pointings enable us to monitor the spectral state of the source. Previous studies of \source have shown that its spectrum can be well described, as is typical for atoll and Z sources, by a soft \texttt{diskbb} component representing disk emission and a \texttt{bbodyrad} component representing Comptonization emission from the spreading and/or boundary layer \citep{DT12}. More detailed modeling of the hard Comptonization component, such as using \texttt{compTT} or \texttt{compTB}, is not possible due to the limited energy bands of \ixpe and \swift. Literature on  \xmm spectra of \source reports wind and reflection features \citep{DT12}, but neither \ixpe nor \swift allows us to study these features. Therefore, in the following analysis, we will apply the simplified spectral model, \texttt{tbabs*(diskbb+bbodyrad)}.

\subsection{Spectroscopy}

Given its superior spectral capabilities, we used \swift data as the primary source for spectral modeling and to monitor spectral variations. The analysis was conducted in the 0.7–8 keV energy band, with best-fit values reported for each observation in Table~\ref{tab:spectra}, including uncertainties at the 68\% confidence level (CL).  Observation IDs are ordered by observation date to illustrate the time evolution of the spectral model.

\begin{table*}[!h]
\caption{Best-fit parameters for the \texttt{TBabs*(diskbb+bbodyrad)} spectral model as obtained by the different \swift spectra.}
\label{tab:spectra}
\begin{tabular}{ccccccccc}
\hline\hline
{ObsID} & \texttt{tbabs} & \multicolumn{2}{c}{\texttt{diskbb}} & \multicolumn{2}{c}{\texttt{bbodyrad}} & $\chi^2$/{d.o.f.} & {Flux}$_{2-8\,\text{keV}}$  & {Hardness} \\
& $N_\text{H}$ (10$^{22}$ \text{cm}$^{-2}$) & $kT_\text{in}$ (keV) & {norm} & $kT_\text{bb}$ ({keV}) & {norm} & & (10$^{-9}$ {erg}~{cm}$^{-2}$~{s}$^{-1}$) & \\
\hline
00036688043 & $4.54^{+0.18}_{-0.14}$ & $1.1^{+0.6}_{-0.2}$ & $250^{+260}_{-160}$ & $1.5^{+0.6}_{-0.1}$ & $100^{+60}_{-90}$ & 0.93 & 6.0 & 0.35 \\
00036688048 & $4.52\pm0.10$ & [1.0] & $290\pm19$ & $1.52\pm0.04$ & $120\pm12$ & 0.91 & 5.9 & 0.37\\
00036688045 & $5.71\pm 0.19$ & $0.67\pm0.05$ & $1100\pm400$ & $1.57\pm0.04$ & $100\pm10$ & 1.13 & 4.1 & 0.45 \\
00036688042 & $5.0\pm0.2$ & $0.70^{+0.10}_{-0.07}$ & $1000^{+700}_{-400}$ & $1.41^{+0.06}_{-0.04}$ & $150\pm30$ & 1.06 & 5.9 & 0.38 \\
00036688047 & $4.62_{-0.18}^{+0.14}$ & $0.88_{-0.10}^{+0.38}$ & $500\pm300$ & $1.38^{+0.18}_{-0.05}$ & $200_{-100}^{+40}$ & 1.12 & 6.0 & 0.36 \\
00036688044 & $4.67_{-0.15}^{+0.17}$ & $0.98_{-0.15}^{+0.39}$ & $400^{+300}_{-200}$ & $1.46^{+0.31}_{-0.09}$ & $130_{-100}^{+60}$ & 1.04 & 5.9 & 0.33 \\
00036688046 & $4.9_{-0.2}^{+0.3}$ & $0.70^{+0.11}_{-0.08}$ & $1100_{-600}^{+900}$ & $1.29_{-0.04}^{+0.06}$ & $230^{+40}_{-50}$ & 1.10 & 5.0 & 0.33 \\
00036688049 & $4.3\pm0.3$ & $1.0_{-0.2}^{+0.6}$ & $320_{-260}^{+630}$ & $1.7_{-0.1}^{+1.3}$ & $80_{-75}^{+65}$ & 1.16 & 5.4 & 0.34 \\
00036688050 & $5.7\pm0.4$ & $0.66^{+0.12}_{-0.09}$ & $1800_{-1000}^{+2000}$ & $1.46_{-0.08}^{+0.11}$ & $130\pm40$ & 0.92 & 4.7 & 0.36 \\
00036688051 & $4.70_{-0.09}^{+0.12}$ & $1.13^{+0.19}_{-0.16}$ & $310_{-120}^{+200}$ & $1.77_{-0.17}^{+0.28}$ & $80_{-40}^{+50}$ & 1.16 & 7.9 & 0.35 \\
\hline
\end{tabular}
\tablefoot{The fits are performed in the energy band 0.7--8.0~keV. Errors correspond to the 68\% CL. The reported flux has a typical uncertainty on the order of $0.3\times10^{-9}$. 
The hardness is defined as in Eq.~\eqref{hr}.}
\end{table*}

In the spectral analysis, the \swift data were fitted using the \texttt{tbabs*(diskbb+bbodyrad)}model. Across the different pointings, the absorption column density was on average $\sim4.7\times10^{22}$~cm$^{-2}$, while the disk temperature ranged between 0.7 and 1.1~keV. Regarding the \texttt{bbodyrad} component -- which in our case is used to describe a Comptonization emission from the boundary layer -- the temperature ranged from $\sim1.3-1.8$~keV. The flux was typically in the range of $\sim$(4--6)$\times10^{-9}$~erg~cm$^{-2}$~s$^{-1}$, except in the final pointing, where it reached the highest value of $\sim8\times10^{-9}$~erg~cm$^{-2}$~s$^{-1}$. The hardness (see Table~\ref{tab:spectra}) defined as
\begin{equation}
\text{Hardness}=\frac{\text{Flux}_{3-8~\text{keV}}-\text{Flux}_{0.7-3~\text{keV}}}{\text{Flux}_{0.7-8~\text{keV}}} 
\label{hr}
\end{equation}
has an almost constant value, except for Observation ID 00036688045, where the flux was minimal and the hardness maximal. However, this \swift observation (third snapshot) was conducted mainly during an \ixpe occultation period. Thus, the \ixpe data are not affected by this slight hardening, and the other snapshots confirm the \ixpe result of an almost constant hardness ratio during the observation. In the following analysis for the spectropolarimetric study, this Observation ID is excluded, and the others are used in the joint fit with the \ixpe data.

\begin{figure}
\centering
\includegraphics[width = 1.\linewidth]{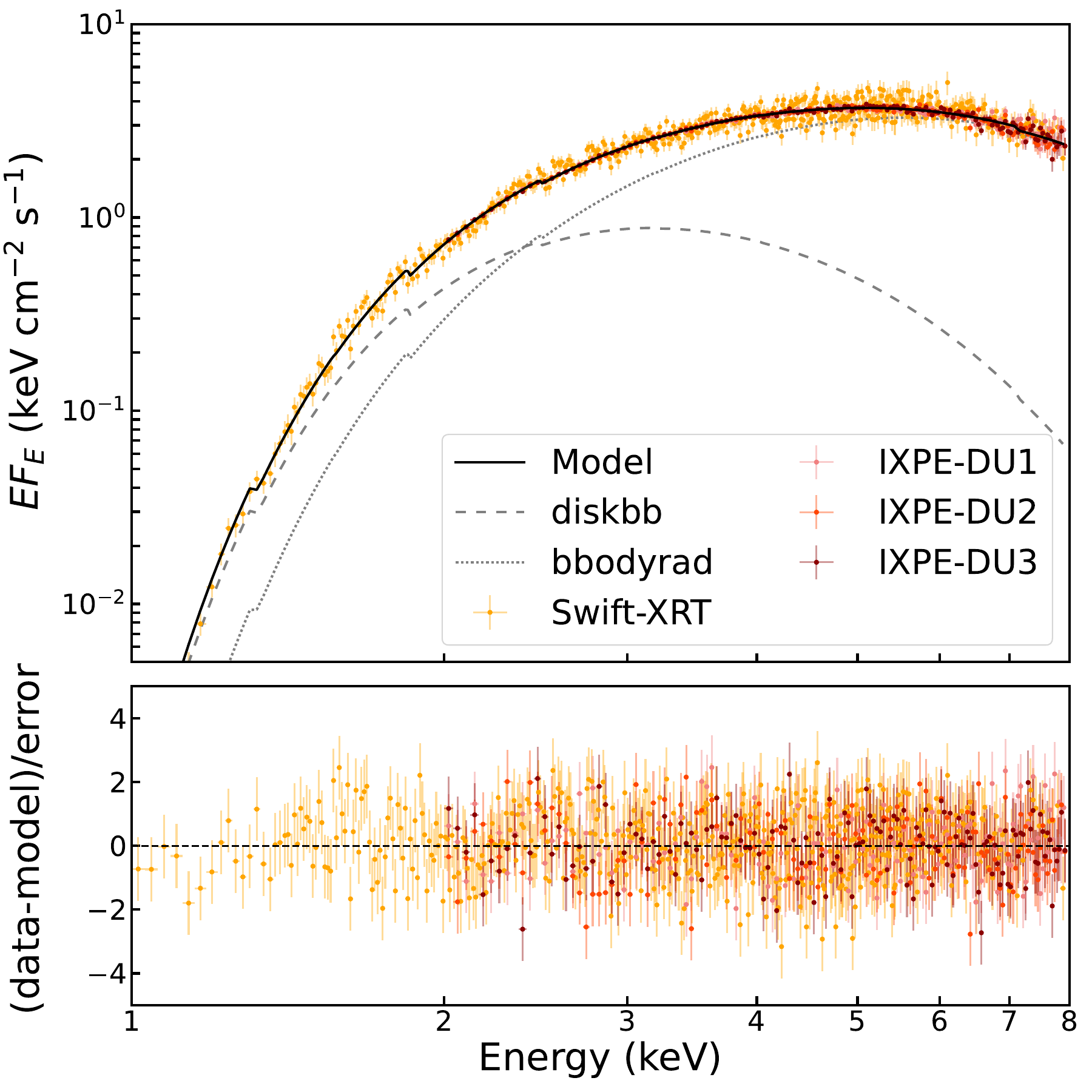}
\caption{Joint spectral fit of \swift and \ixpe data in the 1--8~keV energy range. In the top panel, the spectra of each telescope are reported in $EF_E$ units; residuals are shown in the bottom panel.}
\label{fig:joint_spectra}
\end{figure}

Considering the entire \ixpe observation and the \swift spectra, except the Observation ID 00036688045, we obtain a good joint spectrum model. The best-fit parameters are reported in Table~\ref{tab:joint_spec} and Fig.~\ref{fig:joint_spectra}. To account for the calibration uncertainties in the \ixpe data \citep[see, e.g.,][]{DiMarco_2022b}, we allowed the gain slope and offset of the \ixpe DUs to vary freely.

\begin{table}
\centering
\caption{Best-fit  parameters of the \texttt{const*tbabs*(diskbb+bbodyrad)} model
 for the joint \swift and \ixpe spectra.} 
\label{tab:joint_spec}
\begin{tabular}{ccc}
\hline\hline
Model & {Parameter} & {Value} \\ 
\hline
\texttt{tbabs} & $N_\text{H}$  (10$^{22}$ \text{cm}$^{-2}$) & $4.45_{-0.05}^{+1.00}$ \\ 
\texttt{diskbb} & $kT_{\text{in}}$ (\text{keV}) & $0.88_{-0.07}^{+0.10}$ \\
& \text{norm} & $390^{+140}_{-130}$\\ 
\texttt{bbodyrad} & $kT_{\text{bb}}$ (\text{keV}) & $1.33^{+0.13}_{-0.14}$ \\
& \text{norm} & $228_{-8}^{+20}$\\ 
\texttt{const} & \swift/XRT & [1] \\
& \ixpe-DU1 & $0.722 \pm 0.003$\\
& \ixpe-DU2 & $0.686 \pm 0.003$\\
& \ixpe-DU3 & $0.658 \pm 0.003$\\
$\chi^2$/{d.o.f.} & \multicolumn{2}{c}{991/956=1.04} \\ 
\hline 
\multicolumn{3}{c}{\text{Flux}~($10^{-9}$\,\text{erg}~\text{cm}$^{-2}$~\text{s}$^{-1}$)}  \\
  2--8\text{ keV} & & 6.0 \\
  \texttt{diskbb} & & 1.3 \\
 \texttt{bbodyrad} & & 4.7 \\      
\hline
\end{tabular}
\tablefoot{Errors correspond to the 68\% CL.}
\end{table}

The joint fit confirms the expected result for the average model from several \swift snapshots. In the following spectropolarimetric analysis, the spectral model is fixed to the present one. The \ixpe gain correction is at a level of 0.95 for the slope in all three DUs, with the offset in the range of 70--100 eV.

\subsection{Polarimetry}

\begin{figure}
\centering
\includegraphics[width = 1.\linewidth]{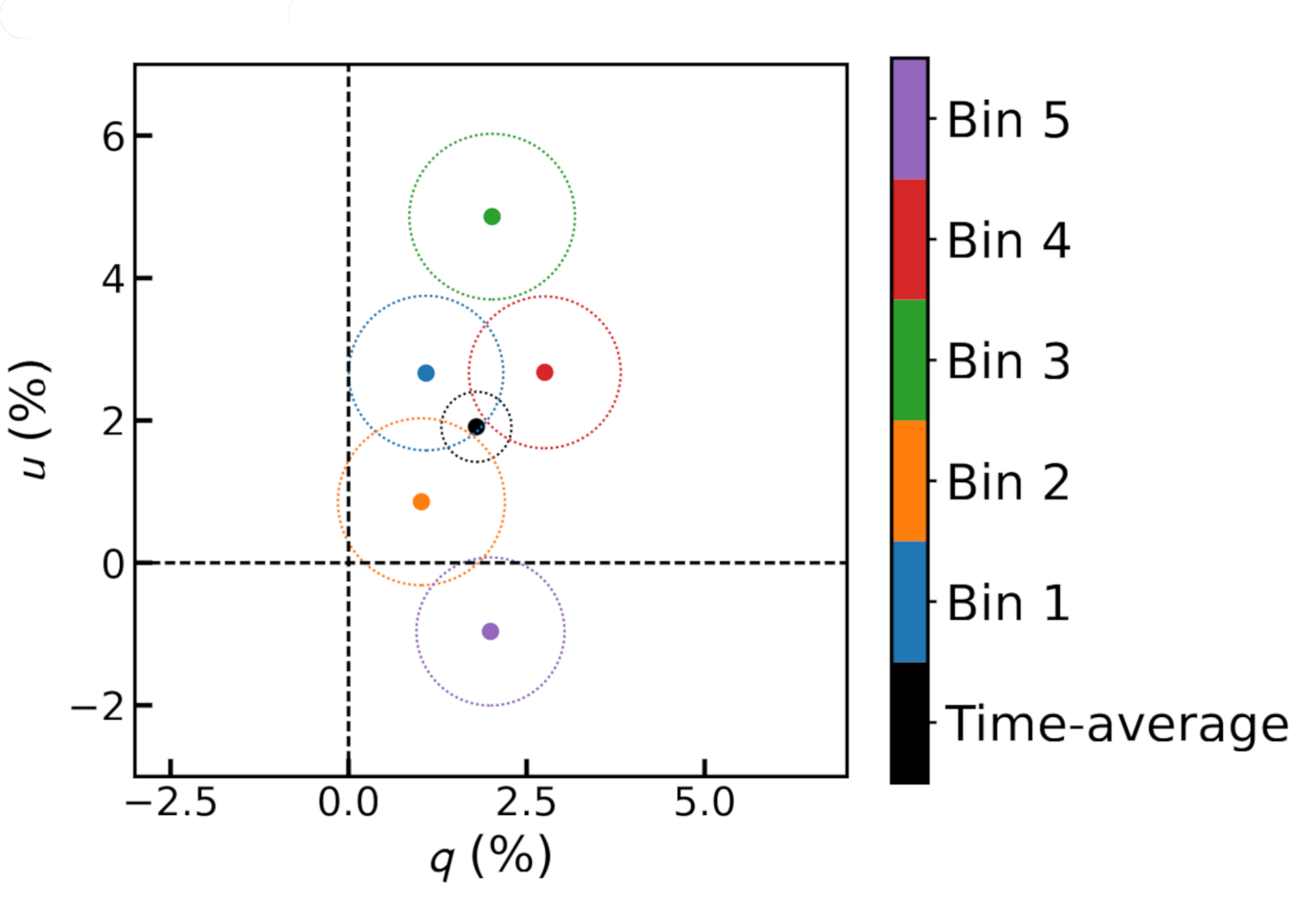}\\
\includegraphics[width = 1.\linewidth]{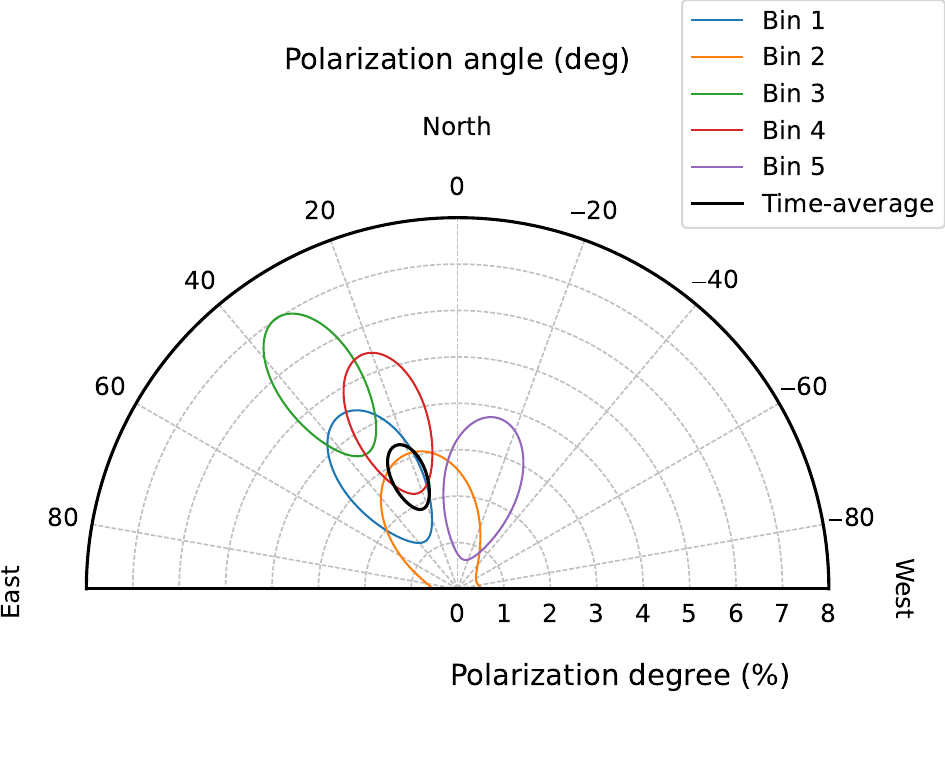}
\caption{Time dependence of the normalized Stokes parameters $q$ and $u$ (top), and of the PD and PA (bottom) obtained by the \texttt{pcube} algorithm using the data separated into 9.5~h time bins. Confidence regions are reported at $68\%$.}
\label{fig:pd_time}
\end{figure}

The spectroscopy reveals that neither the light curve nor the hardness has strong dips or features that would suggest the observation should be separated into several parts to be studied independently. To study the time variability of the polarimetric properties, we followed the same approach as in B24 and split this new \ixpe observation into five equal time bins spanning 9.5~h. The results are reported in Fig.~\ref{fig:pd_time}.

\begin{figure}
\centering
\includegraphics[width = 0.85\linewidth]{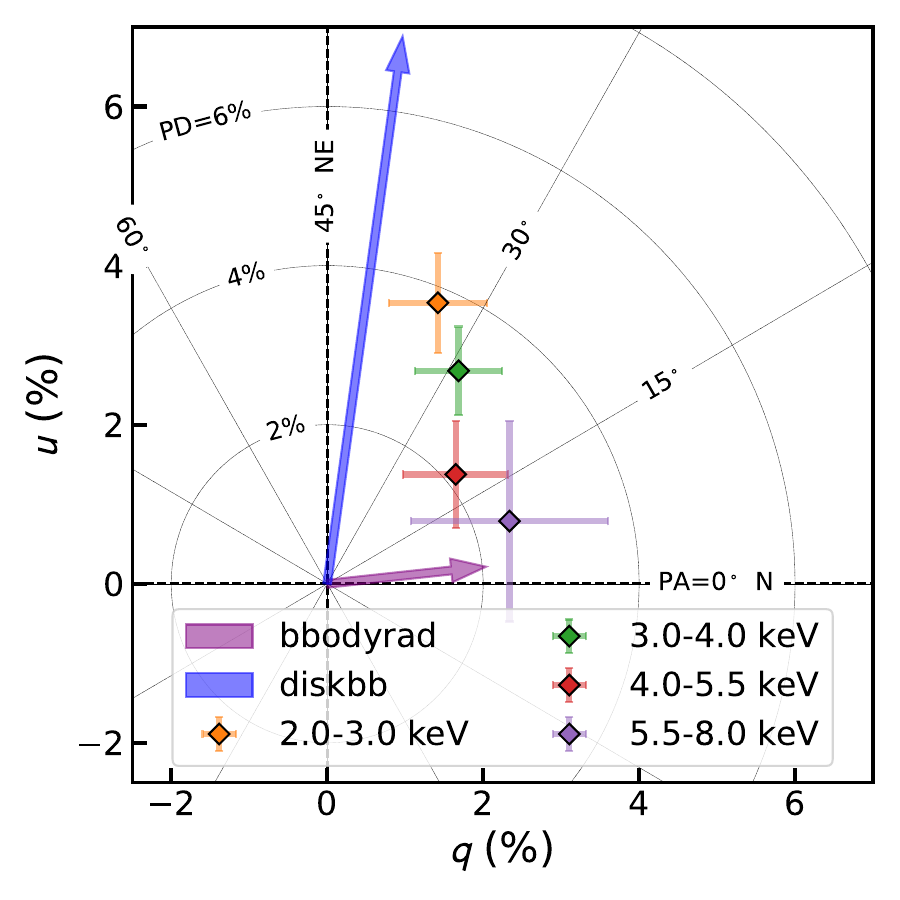}\\
\includegraphics[width = 1.\linewidth]{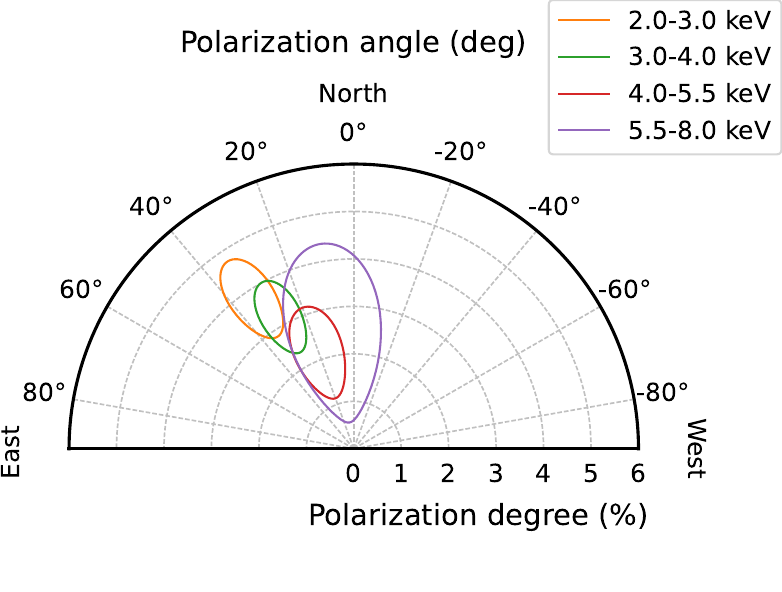}
\caption{Energy-resolved normalized Stokes parameters $q$ and $u$ (top), and PD and PA (bottom) obtained by the \texttt{pcube} algorithm using the reported energy bins for time-averaged data. The confidence regions are reported at $68\%$ CL. The vectors correspond to the best-fit polarization parameters of the \texttt{diskbb} and \texttt{bbodyrad} components, respectively, as obtained using spectropolarimetric analysis and reported in Table~\ref{tab:spec_pol}. }
\label{fig:pol_energy}
\end{figure}

The time dependence of the polarimetric properties shows a hint of variability in PD, completely different from the rotation we observed in the first \ixpe observation. We found it useful to proceed with analyzing the observation as a whole since there are no clear trends. Next, we studied the dependence of the polarimetric properties on energy; the results are reported in Fig.~\ref{fig:pol_energy}. 
The polarization shows a hint of energy dependence, but the significance of this trend is less than 2$\sigma$, as seen later in the spectropolarimetric analysis.
The average in energy and time polarization  during this observation is PD=$2.6\%\pm0.5\%$ with the PA=$23\degr\pm5\degr$ at a significance level of 5.2$\sigma$, corresponding to a secure detection at CL$>$99.9999\%.

\subsection{Spectropolarimetry}

\begin{figure*}
\centering
\includegraphics[width = 0.9\linewidth]{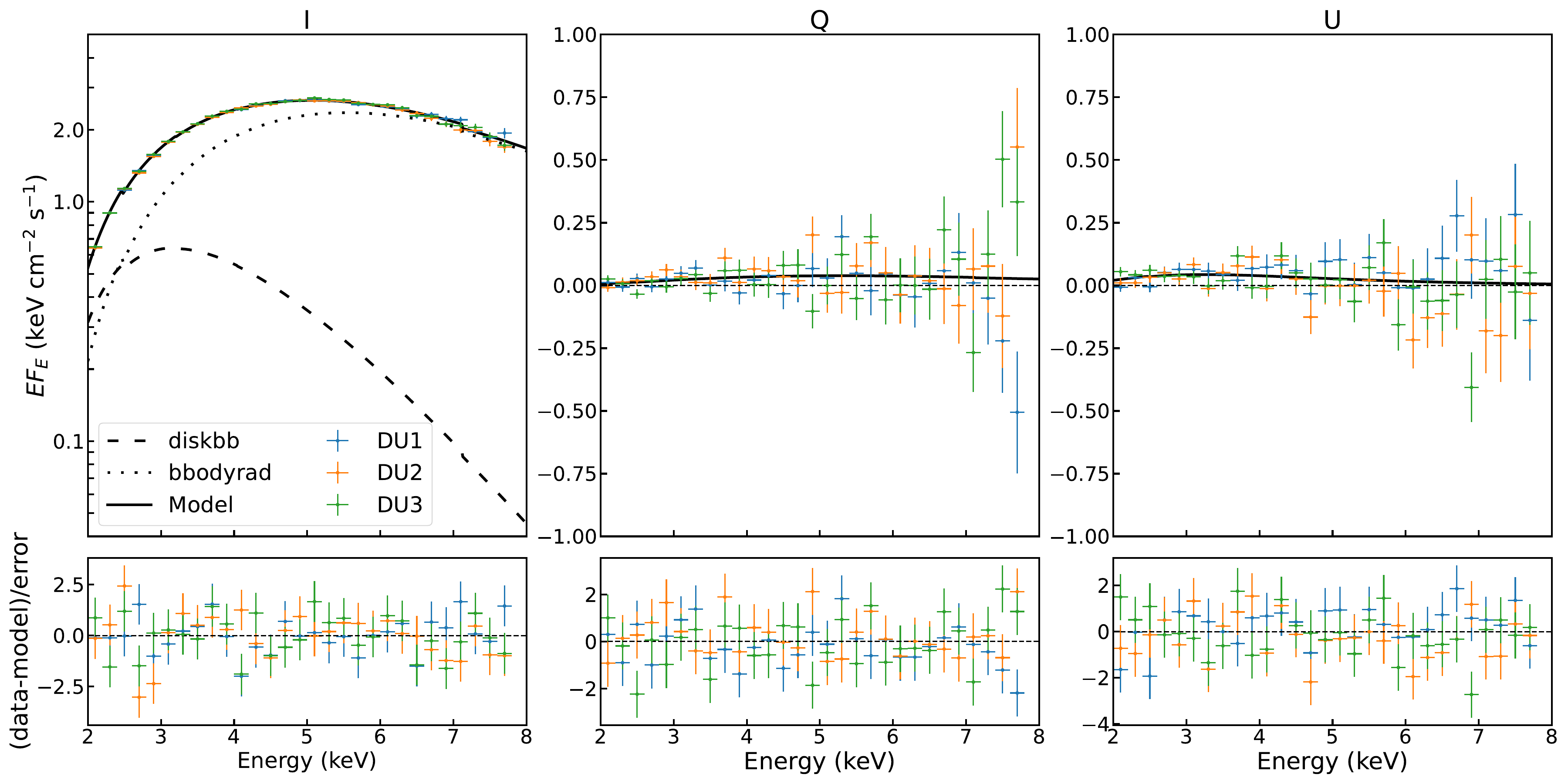}
\caption{Spectral energy distribution of \source in $EF_E$ representation as observed by \ixpe. The left, middle, and right panels are for the Stokes parameters $I$, $Q$, and $U$, respectively. The fit is performed in the 2--8 keV energy band using the three \ixpe detectors and applying the \texttt{tbabs*(diskbb*polconst+bbodyrad*polconst)} model. The total model is shown with the solid black line, while \texttt{diskbb} and \texttt{bbodyrad} are indicated with the dashed and dotted lines, respectively. The lower subpanels show the residuals between the data and the best fit. In the joint spectral fit, we applied a rebinning to obtain at least 50 counts per bin as in \swift, while for the spectropolarimetric analysis, a constant grouping was applied to have energy bins of 200 eV.
}
\label{fig:spec_pol}
\end{figure*}

To study the polarization of the different spectral components, we used the results from the spectral analysis to freeze the spectral model in the spectropolarimetric analysis. First, instead of studying the polarization of each spectral component independently, we applied the \texttt{polconst} model to the entire continuum model to confirm the model-independent analysis. The result is an average PD of $2.4\%\pm0.3\%$ with the PA=$28\degr\pm3\degr$, fully compatible with the model-independent analysis. We then replaced the \texttt{polconst} model with \texttt{pollin} to obtain a quantitative estimation of the dependence of polarization on energy. We obtained nonzero slopes for PA and PD, but they were compatible with zero within a 2$\sigma$ level. Finally, we applied a model with different polarizations for each spectral component, obtaining a PD of $6.1\%\pm1.6\%$ and a PA of $41\degr\pm7\degr$ for the \texttt{diskbb} component, and a PD of $1.6\%\pm0.9\%$ with a PA of $6\degr\pm10\degr$ for the \texttt{bbodyrad} component. The results of the spectropolarimetric analysis are summarized in Table~\ref{tab:spec_pol}; the best-fit plots for the spectropolarimetric analysis are reported for the $I$, $Q$, and $U$ Stokes parameters in Fig.~\ref{fig:spec_pol}.

\begin{table*}
\centering
\caption{\ixpe spectropolarimetric fit of the Stokes parameters $I$, $Q$, and $U$ using the spectral model from Table~\ref{tab:joint_spec}.} 
\label{tab:spec_pol}
\begin{tabular}{cccccc}
\hline\hline
{Model} & {PD}/{$A_1$}~(\%) & $A_{\rm slope}$~(\%\,{keV}$^{-1}$) & {PA}/$\psi_1$~(deg) & $\psi_{\rm slope}$~(deg\,{keV}$^{-1}$) & $\chi^2$/{d.o.f.}\\ \hline
\texttt{tbabs*(diskbb+bbodyrad)*polconst} & $2.4\pm0.3$ & -- & $28 \pm 3$ & -- & 243/250=0.97 \\ 
\hline
 \texttt{tbabs*(diskbb+bbodyrad)*pollin} & $3.8\pm0.8$ & $-0.5\pm$0.3 & $28 \pm 3$ & $-8\pm$4 & 234/248=0.94 \\ 
 \hline
\texttt{tbabs*(diskbb*polconst} & $6.5\pm1.6$ & -- & $41 \pm 7$ & --& 235/248=0.95 \\
\texttt{+bbodyrad*polconst)} & $1.6\pm0.7$ & -- & $3\pm13$ & -- & \\
\hline
\end{tabular}
\tablefoot{Errors correspond to  68\% CL.}
\end{table*}

\section{Discussion}\label{sec:discussion}

\begin{figure}
\centering
\includegraphics[width = 1.\linewidth]{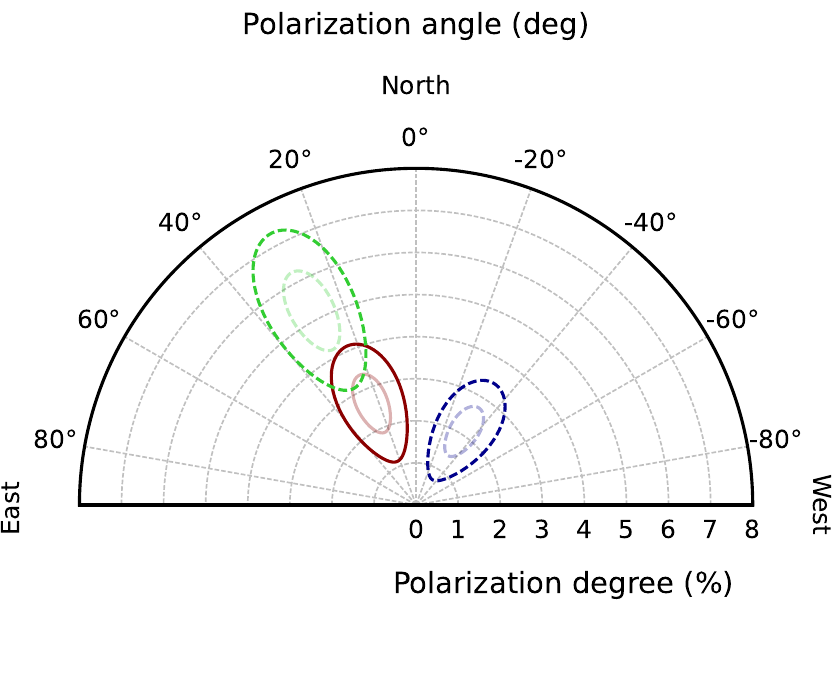}
\caption{Average polarization in the 2--8~keV energy band for the pre-dip (dashed blue lines) and post-dip (dashed green lines) of the first observation of \source, and for the new observation (solid dark red lines). 
Contours correspond to the 68\% and 99\% CLs. A significant detection of polarization at 5.2$\sigma$ is obtained in the new observation with the PA  aligned with the post-dip reported in B24.}
\label{fig:average_pd}
\end{figure}

Although intriguing on its own, the observation reported in this article yields the most insight when compared to the results of the October 2023  observations reported in B24, when we measured an overall PD of $\sim$1.4\% and a PA of $\sim-2\degr$, significantly different from the current results. However, the reason for the depolarization in the first observation was fast variability in the PA. In B24, the observation was separated into the pre-dip, dip, and post-dip states. The results for the dip were poorly constrained, but the other two are presented in Fig.~\ref{fig:average_pd} and compared to the results of the current observation. We note that the current observation is similar in PA to the post-dip state of B24, but the PD is lower probably due to the averaging over time and energy. Bin~3 in Fig.~\ref{fig:pd_time}, for instance, shows a polarization within errors similar to the one measured after the dip in B24. 

Both the HID and the shape of the spectrum confirm that \source was in a soft state throughout the observation. We do not see any significant change of state in the \ixpe and \swift energy bands, although, in the last part of the observation, the flux rises, which is a sign of the source going into an even softer state. If we look at the HID of the source in the October observation (Fig.~2 in B24), we see that the source is almost in the same state, except for the dipping period. During the new observation period, the spectral modeling and decomposition into two main components are better determined, since simultaneous \swift observations offer a better spectral capability than \ixpe, thereby allowing for a better spectropolarimetric analysis. 

The light curve does not show any peculiar variability: unlike the one from the October observation, the current one has no strong dips or rapid changes, and neither does the hardness ratio. Despite this constant hardness, we attempted to study the possible variation in 9.5~h equal time bins similar to B24. This analysis suggests possible variation in PD in Fig.~\ref{fig:pd_time}; in particular, it is interesting to note that in Bin~5, corresponding to a softening of \source, there is lower polarization and an indication of a possible rotation with respect to the other four time bins. The clear rotation of the PA observed during the first observation is not present this time, but the PA appears to have remained aligned with the post-dip polarization measured in B24. 
If we compare the positions of the points on the $(q,u)$ plane with those obtained from the October 2023 observation (Fig.~7 in B24), we see a different pattern; however, the points are still in the I or IV quadrant. The most significant difference lies in the pattern: in October, we observed the points following a straight line path, whereas here, the variability exhibits a different pattern.

Because of the almost constant hardness and flux, we decided to examine the observation as a whole and study the polarization as a function of energy. As presented in Table~\ref{tab:spec_pol}, the slopes of the \texttt{pollin} model are consistent with zero at the 2$\sigma$ level, indicating only a weakly significant dependence of PA and PD on energy. However, it is interesting to note that the spectropolarimetric analysis assigned very different PAs to the two main continuum components. Figure~\ref{fig:pol_energy} (top) illustrates this decomposition: while the PD of the disk is rather high, the contribution of the disk to the total spectrum is low, and it becomes even lower with increasing energy, as the Comptonized component becomes increasingly dominant. The most interesting part here is the unusual difference in PA of the two components of 30\degr--40\degr. This indication can be seen in the energy-resolved analysis, with the 2--3~keV bin near the arrow corresponding to the \textsc{diskbb} component polarization, and the 5.5--8.0~keV energy bin approaching the arrow, which represents the polarization of the \textsc{bbodyrad} component.

From the aligned systems, we expect the polarization vectors of the emission coming from the disk and the Comptonized component to be either nearly parallel or perpendicular to each other. Hence, we expect some misalignment in the system that would account for this difference in the PA. Additionally, it is important to note the exceptionally high PD of the disk component. As stated by \citet{Loktev22}, polarization exceeding 4\% in the \ixpe energy range is not anticipated, even at the highest inclination.
This value is compatible with the spectropolarimetric results at 90\% CL. However, knowing that the source is likely to have an inclination lower than 80\degr \citep{DT12, 2020MNRAS.497.4970T}, we need to add the scattering of the disk emission in the wind to reach a PD of 6.5\%. We assumed that the Comptonized component comes from the boundary layer, as it is expected to align in PA with the disk emission. The observed difference in PAs is less than 45\degr. 

In search of a source of misalignment, we examined previous observations by \ixpe. It is possible, though unexpected, that the rotation axis of the neutron star is slightly misaligned from the rotation axis of the binary system, similar to the \mbox{Cir X-1} \citep{Rankin2024}. In such a scenario, the boundary layer could be dragged out of the disk plane by the motion of the star. Another potential explanation is the peculiar geometry of the wind within the system. 

Finally, we wish to address the significant difference between the two observations of \source. Although the object was in the same spectral state, it was observed half an orbit away from the previous observation, during which the source clearly exhibited a rotation of the PA over time. However, in February 2024, there was no evident monotonic change in the PA, despite ongoing variations. One possibility is that the mechanism responsible for the dip in the light curve during the first observation also caused the rotation of the PA. However, this mechanism remains unidentified. Notably, the recent observation suggests that the PA remained aligned with that observed at the end of the previous observation. As the flux increased and the source transitioned to a softer state at the end of the present \ixpe pointing, the PD decreased and rotated to a PA similar to that observed in October before the dip. This pattern might indicate, as observed in  \mbox{Cir X-1}, a variation in the PA corresponding to the state of the source. This implies a potential variation in the geometry of the hot region across different states. This hypothesis is supported by the October 2023 observation, where the energy dependence of the polarization shifted from an energy-dependent behavior, as seen in \mbox{4U~1820$-$30} \citep{DiMarco23}, to a constant polarization with energy, as observed, for example, in  \mbox{Sco X-1} in \cite{LaMonaca2024}. 
While current data do not allow for a definitive conclusion, we hope that future observations will provide sufficient significance for a detailed analysis of the polarization across different states of atoll and Z sources.

Additionally, the role of the wind remains an open question. Given our lack of knowledge regarding wind behavior during both observations, it is plausible that the wind could contribute to the observed differences in polarimetric properties, although we currently have no means to substantiate this hypothesis. 

\section{Summary}\label{sec:summary}

We analyzed the observation of \source carried out simultaneously by \ixpe and \swift. In this article, we report the highly significant detection of polarization from this source in the soft state. 
Spectroscopic analysis shows that the spectrum is close to that in \citet{DT12}, and the hardness of the emission confirms the lower left banana state, as well as the further softening of the spectrum at the end of this new observation. 
We used a \texttt{diskbb} model for the softer disk emission, and the \texttt{bbodyrad} model to approximate the harder component associated with the spreading or boundary layer. 
The polarimetric analysis provided the overall PD of 2.5\% at >$5\sigma$ CL and a PA of 24\degr. We also see a marginally significant decrease in PD and PA with energy. The same trend was confirmed by the spectropolarimetric analysis, which showed that the softer component was more strongly polarized than the harder component, with the PA differing by $\sim$40\degr. 
The results of the spectropolarimetric analysis using \textsc{xspec} performed under the assumption of one overall polarization model were in agreement with the \texttt{pcube} polarimetric results. 

We studied the differences and similarities between the first (B24) and the second observations of \source by \ixpe. Our main conclusion is that the overall polarization is significantly higher in the second observation despite the source being in the same state during the two observations. Depolarization was caused by the variability in polarimetric properties, which was larger during the first observation. We have discussed possible reasons for the different polarimetric behavior in the two observations.

\begin{acknowledgements}

This research used data products provided by the IXPE Team (MSFC, SSDC, INAF, and INFN) and distributed with additional software tools by the High-Energy Astrophysics Science Archive Research Center (HEASARC), at NASA Goddard Space Flight Center (GSFC). The Imaging X-ray Polarimetry Explorer (IXPE) is a joint US and Italian mission.  
The Italian contribution is supported by the Italian Space Agency (Agenzia Spaziale Italiana, ASI) through contract ASI-OHBI-2022-13-I.0, agreements ASI-INAF-2022-19-HH.0 and ASI-INFN-2017.13-H0, and its Space Science Data Center (SSDC) with agreements ASI-INAF-2022-14-HH.0 and ASI-INFN 2021-43-HH.0, and by the Istituto Nazionale di Astrofisica (INAF) and the Istituto Nazionale di Fisica Nucleare (INFN) in Italy. 
We thank the \textit{Swift} Project Scientists for approving our DDT request to observe \source. 
This research has been supported by the Finnish Cultural Foundation grant 00240328 (AB) and the Academy of Finland grant 333112  (AB, JP, SVF,  VL).
ADM and FLM are partially supported by MAECI with grant CN24GR08 “GRBAXP: Guangxi-Rome Bilateral Agreement for X-ray Polarimetry in Astrophysics”.
\end{acknowledgements}

%
%
\bibliographystyle{aa}
\bibliography{biblio}

\end{document}